\documentclass[
    ,final            
  ]
  {aipproc}

\newcommand{\be}[1]{\begin{equation} #1 \end{equation}}

\newcommand{\paren}[1]{\left( #1 \right)}

\newcommand{\bc}{\begin{center}}
\newcommand{\ec}{\end{center}}

\def\eg{{\it e.g.\ }}

\def\ie{{\it i.e.\ }}
\def\cf{{\it cf.\ }}

\layoutstyle{6x9}

\begin{document}

\title{Information geometry of asymptotically AdS black holes}



\author{Jan E. \AA man}{
  address={Department of Physics, Stockholm University, 106 91 Stockholm,  Sweden}
}

\author{Narit Pidokrajt}{
  address={Department of Physics, Stockholm University, 106 91 Stockholm,  Sweden}
}

\author{John Ward}{
  address={Department of Physics and Astronomy, University of Victoria, Victoria, BC, V8P 1A1, Canada}
}

\begin{abstract}
We investigate thermodynamic geometries of two families of asymptotically Anti-de Sitter black holes, \ie the Reissner-Nordstr\"om Anti-de Sitter in four dimensions and the BTZ black hole. It is found that the Anti-de Sitter  space renders the geometry nontrivial (\cf the Reissner-Nordstr\"om black hole in asymptotically flat background). The BTZ black hole's thermodynamic geometry is trivial despite the fact that it is characterized by the (negative) cosmological constant. As a matter of curiosity we compute thermodynamic geometry of these black holes regarding the cosmological constant as a true parameter but no physically significant results can be derived.

\end{abstract}

\maketitle


In this article we discuss the use of information geometry~\cite{Johnston:2003ed} to study asymptotically Anti-de Sitter (AdS) black holes (BHs)\footnote{This family of BH was studied \eg in~\cite{Aman:2003ug}, \cite{Shen:2005nu} and \cite{Quevedo:2008xn}. In \cite{Quevedo:2008xn} the authors employ a modified thermodynamic geometry according to the so-called geometrothermodynamic principle~\cite{Quevedo:2007mj, Alvarez:2008wa}, \ie the thermodynamic metric in question should be invariant under Legendre transformations.}. Information geometry in the BH physics context is known as the Ruppeiner geometry~\cite{ruppeiner79} or thermodynamic geometry whose metric is defined as a Hessian of the entropy function on the state space of the BH in question. This geometric method serves as an alternative route to the study of BH thermodynamics via Riemannian geometry. In the last five years it has been applied to a number of BH families. Some well-known results \cite{Aman:2003ug}  are as follows: the Ruppeiner metric is flat for the BTZ, 4D dilaton~\cite{Aman:2007ae} and Myers-Perry (MP) Reissner-Nordstr\"om (RN) BHs~\cite{Aman:2005xk} while curvature singularities occur for the RN-AdS and MP Kerr BHs. The information metric of thermodynamics is defined as
\begin{equation}
g^{R}_{ij} = -\partial_i \partial_j S(X)
\end{equation}
where $X = (S, N^a)$, $S$ is the system's entropy and $N^a$ are other extensive (mechanically conserved) parameters such as mass and angular momentum in the case of BH systems. The minus sign arises because entropy is a concave function. Interpretations of the geometries associated with the metric are discussed in \cite{Ruppeiner:1995zz} and references therein. There have been a number of results indicating that this geometry measures the underlying microscopic physics. In particular for systems with no statistical mechanical interactions (\eg the ideal gas), the Ruppeiner geometry is flat and vice versa. The Ruppeiner metric is conformally related to the so-called  Weinhold metric defined in its original coordinates as
\begin{equation}
g^W_{ij} = \partial_i \partial_j U(X).
\end{equation}
The function $U(X)$ is the system's internal energy (mass for the BHs). The two metrics are related via $g^{R}_{ij} = \frac{1}{T}g^W_{ij}$ where $T$ is the temperature of the system. However interpretation of its curvature is somewhat vague. The Weinhold metric can be very useful for some systems where the metric is complicated in original Ruppeiner coordinates. As a matter of fact, fairly frequently one first attempts the calculations in Weinhold coordinates and then conformally transforms it into Ruppeiner metric, yet in Weinhold coordinates. We have learned that the metric signature reflects the sign of specific heat of the system, \ie it is Euclidean if the BH has positive specific heat like that for the BTZ, whilst Lorentzian for MP RN, RN-AdS, MP Kerr BHs. For 2D dilaton BH the Ruppeiner metric can have either Euclidean or Lorentzian signature depending on the parameter of the model \cite{Aman:2006mn}. 

In this brief article we discuss results from the study of the AdS BHs because they are systems of interest in many regards. For example, the non-extremal AdS BHs have found prominence within the AdS/CFT correspondence \cite{Maldacena:1997re} which relates strongly coupled conformal field theories (CFT) in $D$ dimensions to weakly coupled gravity in $D+1$ dimensions. Therefore the Ruppeiner geometry of 4D AdS BH could possibly shed light on field theory interactions in three-dimensions. With this in mind, we consider charged BH solutions setting $J=0$, since charged BH solutions have found considerable application in the literature.  

\medskip
{\bf Reissner-Nordstr\"om-AdS black holes} 
\medskip

The Hawking temperature of RN-AdS is given by $T_H = \frac{1}{4\sqrt{S}}\left(1-\frac{Q^2}{S} + \frac{3S}{l^2}  \right)\ $. In our previous work \cite{Aman:2003ug} the Ruppeiner metric was calculated to be
\begin{equation}
\label{bh1}
ds^2_W=\frac{1}{8S^{3/2}}\left[-\left(1-\frac{3Q^2}{S}-\frac{3S}{l^2}\right)dS^2
-8QdSdQ+8SdQ^2\right].
\end{equation}
The Weinhold curvature is a function which vanishes in the extremal limit, and the Ruppeiner curvature is
\begin{eqnarray}
\label{4}
\small
R_R &=& \frac{9}{l^2}\frac{\left(\frac{Q^2}{S}+\frac{3S}{l^2}\right)\left(1-\frac{Q^2}{S}-\frac{S}{l^2}\right)}{\left(1
-\frac{Q^2}{S}-\frac{3S}{l^2}\right)^2\left(1-\frac{Q^2}{S}+\frac{3S}{l^2}\right)}.
\end{eqnarray}
We observe that the Ruppeiner curvature diverges both in the extremal limit, ${\frac{Q^2}{S} = 1 + \frac{3S}{l^2} = 1 - \Lambda S}$, and along the curve where the metric changes signature, which is where the thermodynamical instability sets in (where the metric (\ref{bh1}) changes signature).  
 
\medskip
{\bf BTZ black hole} 
\medskip

The BTZ BH~\cite{Banados:1992wn} has a flat Ruppeiner geometry like that of the RN BH, whilst its Weinhold geometry is curved. The metric signature is Euclidean. We investigate the information geometry of this BH by starting with the Weinhold metric as it is simpler. The Smarr mass is $M = S^2 + {J^2}/{(4S^2)}$.  The BH's Hawking temperature is given by $T = 2S - {J^2}/{(2S^3)}$ whereas its angular velocity takes a rather simple form $\Omega = {J}/{(2S^2)}$. The heat capacity is easily calculated to be $C = {S(4S^4 - J^2)}/{(4S^4 + 3J^2)}.$ We can readily seen that the heat capacity of the BTZ BH is positive, hence it is thermodynamically stable. The Weinhold metric for the BTZ BH is a nonflat. The Ruppeiner metric in a diagonal form reads
\be{
\small
ds^2_R = \frac{1}{S}dS^2 + \paren{\frac{S}{1 - u^2}}du^2,
\label{eq:BTZ-Ruppeiner}
}
where we have used $u = \frac{J}{2S^2}; \, u\in [-1,1]$. The Ruppeiner metric in (\ref{eq:BTZ-Ruppeiner}) is flat, in other words the space of its thermodynamic state is a flat space. It can be depicted as a wedge of a Euclidean flat space in polar coordinates.

\begin{figure}[h]
\caption{The state space for RN-AdS BHs; the coordinates are $u$ and $S$ and the cosmological constant decreases as we go from A to C. The shaded region has a Lorenzian metric.}
\includegraphics[scale=.45, angle=90]{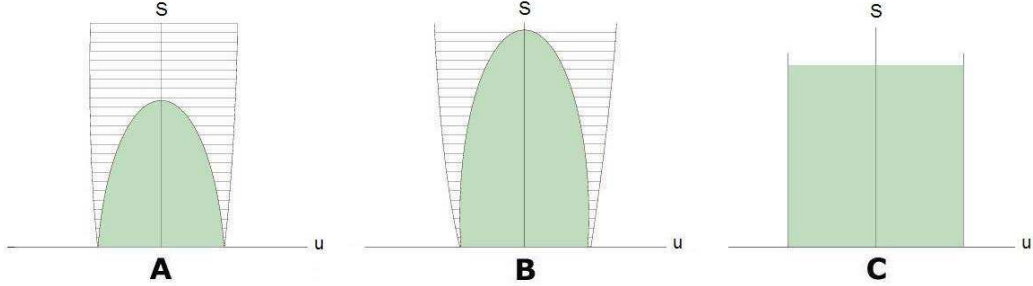}
\end{figure} 
\medskip
{\bf When $\Lambda$ is regarded as a parameter}
\medskip

Previously the study of information geometry of AdS BHs was carried out with BH's extensive parameters \eg mass, spin and charge, meaning that the entropy would be a function of those parameters.  Out of curiosity we wish to compare the Ruppeiner and Weinhold geometries of the associated phase spaces for cases where the cosmological constant is regarded as a true constant and those for which it is regarded as a parameter of the theory. Thus   $S = S(M, Q, J, \Lambda)$ but we will limit our investigation to 3D state space due to mathematical complications that arise when one goes beyond 3D. For 4D BHs in AdS space we will use a Smarr mass formula given in~\cite{Caldarelli:1999xj} with simplication 
\begin{equation}
\small
M = \frac{\sqrt{S}}{2} \sqrt{ \paren{1  -  \frac{\Lambda S}{3} + \frac{Q^2}{S}}^2 + \frac{4J^2}{S^2}\paren{1 - \frac{\Lambda S}{3}}}
\end{equation}
 where we have used $l^2 = -3/\Lambda$ and $\Lambda$ is the negative cosmological constant. This formula represents the mass of the Kerr-Newman-AdS BH, which is the most general BH solution in pure Einstein gravity in 4D.  For the RN-AdS BH we have the Weinhold metric in the following form
\begin{eqnarray}
\small
\label{7}
ds_W^2=\frac{1}{8S^{3/2}}\left[-\left(1-\frac{3Q^2}{S}+\Lambda S\right)dS^2 
-8QdSdQ+8SdQ^2-4S^2dSd\Lambda\right]\ ,
\end{eqnarray}
whose curvature is zero whereas its counterpart, the Ruppeiner metric is nonflat with the curvature 
\begin{eqnarray}
\small
R_R &=& \frac{1}{4S}{\left(7-17\frac{Q^2}{S}+3\Lambda S\right)}{\left(1-\frac{Q^2}{S}-\Lambda S\right)^{-1}}\ .
\end{eqnarray}
It is readily seen that the Ruppeiner curvature blows up in the extremal limit. The BTZ BH involves calculations in 3D phase space, \ie the Ruppeiner metric becomes a function of $(M, J, l)$. We start from the expression of the entropy 
\be{
S = l\sqrt{\frac{M}{2}}\left[1+\sqrt{\left(1-\frac{J^2}{M^2l^2}\right)} \;\right]^{1/2} .
}
The Ruppeiner metric is too lengthy to display here, and its curvature scalar is nonflat as shown below
\begin{equation}
R_R = \frac{13 -7\frac{J^2}{M^2 l^2} -8\rho}{l \sqrt{2M} \rho (\rho -2) \sqrt{1+\rho}}
\end{equation}
where $\rho = \sqrt{1-\frac{1}{M^2 l^2}}$. The curvature blows up at $\rho = 2$, more precisely at $l^2 = \frac{-1}{3M^2}$ which is not a physically significant point as far as we know about the BTZ BH. Surprisingly the Weinhold metric is flat.

\begin{theacknowledgments} 

{Narit Pidokrajt is supported by Doktorandtj\"anst of Stockholm University and would like to thank the ILIAS section of the Gravitational Waves of the Astroparticle network for financial support for his participation in the ERE2008 meeting in Salamanca. NP would also like to acknowledge the Helge Ax:son Johnsons stiftelse for a scholarship for his project on black hole information geometry. }

\end{theacknowledgments}

\end{document}